\documentclass[aps,prb,twocolumn,amsmath,superscriptaddress,groupedaddress]{revtex4}  
\usepackage{graphicx}  
\usepackage{dcolumn}   
\usepackage{bm}        
\usepackage{amssymb}   

\begin{document}

\title{Coulomb impurity under magnetic field in graphene: a semiclassical approach}

\author{Yuhui Zhang}
\affiliation{ National High Magnetic Field Laboratory and
  Department of Physics, Florida State University, Tallahassee, FL
  32306, USA }
\author{Yafis Barlas}
\affiliation{ National High Magnetic Field Laboratory and
  Department of Physics, Florida State University, Tallahassee, FL
  32306, USA }
\author{Kun Yang}
\affiliation{ National High Magnetic Field Laboratory and
  Department of Physics, Florida State University, Tallahassee, FL
  32306, USA }

\date{\today}

\begin{abstract}
We address the problem of a Coulomb impurity in graphene in the presence of a perpendicular uniform
magnetic field.  We show that the problem can be solved below the supercritical impurity magnitude within the WKB approximation.  Without impurity the semiclassical energies correctly reproduce the Landau level spectrum. For a given Landau level the WKB energy depends on the absolute value of angular momentum in a way which is consistent with the exact diagonalization result. Below the supercritical impurity magnitude, the WKB solution can be expanded as a convergent series in powers of the effective fine structure constant. Relevance of our results to validity of the widely used Landau level projection approximation is discussed.
\end{abstract}

\maketitle

\section{INTRODUCTION}
Graphene, a two-dimensional honeycomb lattice of carbon atoms,\cite{Geim,Castro Neto} at low energies can be described by massless Dirac fermions.\cite{Novoselov,ZhangI} This is evident in graphene's Landau level structure which leads to anamolous integer quantum Hall effect (QHE) with plateaus at $\sigma_{xy} = 4(n+1/2)e^2/h$ $(n= 0,\pm 1,\pm2, \cdots)$.\cite{Novoselov,ZhangI} As sample qualities improved experiments began to reveal a large number of additional Hall plateaus~\cite{ZhangII} {\em not} expected from Landau quantization alone;
the gaps associated with these plateaus could only be induced by electron-electron interactions.\cite{kunreview} More recently the importance of strong electron-electron interactions was firmly established after fractional quantum Hall effects (FQHE) were revealed by transport measurements on suspended graphene samples~\cite{graphene_fqhe1,graphene_fqhe2} and graphene on hexagonal Boron Nitride~\cite{BNsubstrate} (h-BN) substrates.\cite{graphenefqhreview}

Current theories of FQHE in conventional semiconducting systems rely heavily on the notion of the projection of interactions onto a single Landau level (i.e. in general cases Landau level mixing is neglected). The appropriateness of the Landau level projection is not trivially obvious for the case of graphene. In semiconducting 2D electron gas this is formally achieved in the limit of a large magnetic field $B \to \infty$. This is justified because Coulomb interaction $V_{e-e} \sim e^2/\varepsilon l_{B}$ scales as $\sqrt{B}$ while the single particle Landau level gap $\hbar \omega_{c}$ scales as $B$. So it can be argued there that for large magnetic fields, interactions between electrons in the lowest partially-filled Landau level cannot induce transitions to the higher Landau levels. The projection of the interactions onto the lowest partially-filled Landau level is not as clearly justified in graphene because the single particle Landau level gaps in graphene also scale as $ \sim \sqrt{B}$, which is the same as the interaction strength. Most FQHE theories for graphene have nevertheless assumed that Landau level mixing is an inessential complication that can be ignored.\cite{graphenefqhreview}

For this reason, it is important to study massless Dirac fermions in the presence of Coulomb interaction and a quantizing magnetic field, and validate Landau level projection approximation. If Landau level projection is a valid approximation, effects of Landau level mixing can be treated perturbatively.
The simplest case would be a two body problem. For non-relativistic particles with Galilean invariance, a two body problem is equivalent to a one-body problem once we separate the center-of-mass and relative motions.
However such a separation is not possible for Dirac fermions, and the two body problem cannot be solved analytically.
In the absence of magnetic field solutions of two-body problems with zero center of mass momentum are possible,\cite{twobody} but these solutions do {\em not} generalize to the present case with magnetic field.
Therefore as a first step we study instead a massless Dirac fermion in the presence of a Coulomb impurity and a uniform magnetic field in this work, and address the following question: Can Landau level mixing effects induced by the Coulomb impurity be treated perturbatively or not? In addition to the interest in its own right, we note many features in this one-body problem such as non-linear screening and supercritical instabilities have direct generalizations in the many body case, for example exciton condensation or spontaneous mass generation.\cite{Kotov}

Unlike the Coulomb impurity problem\cite{Pereira,Shytov} in the {\em zero} magnetic field case, this problem can not be solved exactly in closed analytic form when a uniform magnetic field is present. We instead apply the semi-classical WKB method to solve this problem. While approximate, the WKB method is {\em non-perturbative} in the potential. As we will show, it gives rise to the {\em exact} Landau level spectrum in the absence of the Coulomb impurity, and numerically very accurate energy spectra in its presence for most cases; the latter is established by comparing with exact diagonalization calculation using a truncated Hilbert space that keeps a very large number of Landau levels. By expanding the WKB solutions in power series of Coulomb impurity strength, we show that the series is convergent as long as the Coulomb impurity strength is below the supercritical instability critical point(to be discussed in more detail later), thus establishing the perturbative nature of the Coulomb potential induced Landau level mixing effects. Our results thus lend support to the Landau level projection approximation in this limited parameter range.

We would like to stress the importance of supercritical instabilities in the Coulomb impurity problem in a magnetic field. Without the magnetic field, supercritical instabilities have been investigated by many authors.\cite{GamayunII,Pereira,Shytov} For massless Dirac fermions, it is accompanied with an infinite number of quasi-localized resonances in the hole sector. When the magnetic field is added, these supercritical instabilities are also present at the same critcal value of the impurity strength $g_{c}$. This is because supercritical instabilities are only determined by the short range behavior of the effective potential. The contribution to the effective potential induced by the presence of a quantizing magnetic field vanishes as $ r \to 0$, hence it does not influence the short-distance part of the effective potential.
Beyond $g_{c}$, each Landau level mixes with the quasistationary levels and the whole Hilbert space can not be truncated into a single Landau level. Below ${g_{c}}$, we can use the WKB approximation to solve for the wavefunctions and energy spectrum of a massless Dirac fermion. This method also captures the characteristic features of the supercritical instabilities. Below ${g_{c}}$ it has discrete energy level solutions, which becomes continuous beyond the critical point ${g_{c}}$.
This signals the {\em breakdown} of Landau level projection.

In Sec. II, we outline the WKB method for 2D massless Dirac particles in a uniform magnetic field, and obtain the WKB wavefunctions and Bohr-Sommerfeld (BS) quantization condition for eigenenergies. The BS condition is compared with its counterpart of Schrodinger particle. In Sec. III, the WKB results are shown for cases with and without Coulomb impurity. We also compare the semiclassical energies with energies obtained from exact diagonalization. Sec. IV addresses the convergence of the WKB energies when expanded in powers of Coulomb impurity strength. Finally, we provide a detailed derivation (using Zwaan's method) of BS condition in the Appendix.

\section{WKB METHOD FOR GRAPHENE}
\subsection{Outline of the Problem}
Consider the problem of a single Coulomb impurity in a homogeneous magnetic field perpendicular to the plane of graphene. Define ${H_{K(K')}^{+(-)}}$ as the Hamiltonian of the problem with positive (negative) Coulomb impurity at the ${K}$ ${(K')}$ point of Brillouin zone. With negative Coulomb impurity, close to the ${K}$ point, the electron quasiparticle states are described by the Dirac Hamiltonian

\begin{equation}H_{K}^{-}=\hbar v_{F}(\frac{1}{\hbar}\boldsymbol{\sigma}\cdot\boldsymbol{\Pi}+\frac{g}{r}),\end{equation}
where ${v_{F}\approx 10^{6}m/s}$ is the Fermi velocity, the canonical momentum ${\boldsymbol{\Pi}=-i\hbar\boldsymbol{\nabla}+(e/c)\boldsymbol{A}}$ includes the vector potential ${\boldsymbol{A}}$ corresponding to the magnetic field, ${\sigma_{i}}$ are the Pauli matrices,
${g=Z\alpha}$ in which ${Z}$ is the impurity charge, $\alpha=e^{2}/(\kappa\hbar v_{F})$
is graphene's fine structure constant, and ${\kappa}$ is the effective dielectric constant. (1) does not involve inter-valley scattering, because in Fourier space Coulomb potential behaves like ${1/q}$ and is dominated by small ${q}$. For conventional ${SiO_{2}}$ substrates ${\kappa\approx2.4}$, giving ${\alpha\approx0.92}$ which is much larger than that in QED (${\alpha\approx 1/137}$).

We use the symmetric gauge ${(A_{x},A_{y})=(B/2)(y,-x)}$. Resorting to the rotational symmetry of the system, the eigenfunctions can be written in cylindrical coordinates as

\begin{equation}
\Psi_{l}(r,\phi)=\frac{1}{\sqrt{r}}\left(\begin{array}{c}
F(r)e^{i(l-1)\phi}\\
iG(r)e^{il\phi}\end{array}\right),
\end{equation}
and the radial eigenequation reads
\begin{equation}
\begin{split}
&\left(\begin{array}{cc}
\frac{g}{r}-\frac{\epsilon}{l_{B}} & (\partial_{r}+\frac{l-1/2}{r}-\frac{1}{2l_{B}^{2}}r)\\
(-\partial_{r}+\frac{l-1/2}{r}-\frac{1}{2l_{B}^{2}}r) & \frac{g}{r}-\frac{\epsilon}{l_{B}}\end{array}\right)\\
&\times\left(\begin{array}{c}
F(r)\\
G(r)\end{array}\right)=0,
\end{split}
\end{equation}
where ${l_{B}=\sqrt{\hbar c/(eB)}}$ is the magnetic length, ${\epsilon=E l_{B}/(\hbar v_{F})}$ in which ${E}$ is the eigenenergy of the Hamiltonian (1) and ${\epsilon}$ is dimensionless, ${l=0}$, ${\pm1}$, ${\pm2}$, ${\ldots}$ is the orbital angular momentum quantum number.

Different signs of Coulomb impurity can be related by the operation
\begin{equation}
\sigma_{z}H_{K(K')}^{\pm}\sigma_{z}=-H_{K(K')}^{\mp}.
\end{equation}
It implies that, in a certain valley, a solution ${\left|\Psi\right\rangle }$ to the Dirac equation with energy ${E}$ for positive (negative) Coulomb impurity, has a conjugate partner ${\sigma_{z}\left|\Psi\right\rangle}$ with energy ${-E}$ for negative (positive) Coulomb impurity. On the other hand, different valleys can be related by the operation
\begin{equation}
\sigma_{x}H_{K(K')}^{\pm}\sigma_{x}=H_{K'(K)}^{\pm}.
\end{equation}
Hence with the same Coulomb impurity, a solution ${\left|\Psi\right\rangle }$ to the Dirac equation with energy ${E}$ in valley ${K}$ ${(K')}$, has a conjugate partner ${\sigma_{x}\left|\Psi\right\rangle}$ with energy ${E}$ in valley ${K'}$ ${(K)}$. Therefore, it is enough to solve the problem of negative Coulomb impurity at the ${K}$ point.

Write

\begin{equation}
\left(
\begin{array}{c}
 F(r) \\
 G(r)
\end{array}
\right)=(\frac{\epsilon}{l_{B}} -\frac{g}{r})^{\frac{1}{2}} \left(
\begin{array}{c}
 u (r) \\
 v (r)
\end{array}
\right),
\end{equation}
Eq. (3) can be written as two Schrodinger-like equations,
\begin{equation}
\begin{split}
-u''(r)+U_{1}(r) u (r)=\frac{\epsilon ^2}{l_{B}^2} u (r),\\
-v''(r)+U_{2}(r) v (r)=\frac{\epsilon ^2}{l_{B}^2} v (r),
\end{split}
\end{equation}
where

\begin{equation}
\begin{split}
U_{1}(r)=&\frac{j^2-j-g^2}{r^2}+\frac{g (1-j)}{r^3 \left(\frac{\epsilon}{l_{B}} -\frac{g}{r}\right)}+\frac{3 g^2}{4 r^4 \left(\frac{\epsilon}{l_{B}} -\frac{g}{r}\right)^2}\\
&+\frac{2g\epsilon}{l_{B}r}+\frac{r^{2}}{4l_{B}^{4}}-\frac{j+\frac{1}{2}}{l_{B}^{2}}+\frac{g}{2l_{B}^{2}r\left(\frac{\epsilon}{l_{B}}-\frac{g}{r}\right)},
\end{split}
\end{equation}
\begin{equation}
\begin{split}
U_{2}(r)=&\frac{j^2+j-g^2}{r^2}+\frac{g(1+j)}{r^3 \left(\frac{\epsilon}{l_{B}}-\frac{g}{r}\right)}+\frac{3 g^2}{4 r^4 \left(\frac{\epsilon}{l_{B}} -\frac{g}{r}\right)^2}\\
&+\frac{2g\epsilon}{l_{B}r}+\frac{r^{2}}{4l_{B}^{4}}-\frac{j-\frac{1}{2}}{l_{B}^{2}}-\frac{g}{2l_{B}^{2}r\left(\frac{\epsilon}{l_{B}}-\frac{g}{r}\right)},
\end{split}
\end{equation}
and ${j=l-1/2}$ is the total angular momentum quantum number. Although we have (seemingly decoupled) Schrodinger-like equations (7), ${u(r)}$ and ${v(r)}$ are still related to each other. The reason is that the final wavefunction (2) is a spinor, which is the superposition of the states in sublattice A and B (corresponding to ${u(r)}$ and ${v(r)}$ respectively). The ratio of the two functions ${u(r)}$ and ${v(r)}$ is determined by Eq. (3).

Morse and Feshbach\cite{unsolvable} classified the solutions of second-order ordinary differential equations by types of singular points of the equations. With two regular (${r=0}$, ${gl_{B}/\epsilon}$) and one irregular (${r\rightarrow \infty}$) singular points respectively, Eq. (7)'s exact solutions can not be expressed in closed form in terms of known special functions. At short distance limit ${U_{1,2}(r)\to(j^{2}-g^{2}-1/4)/r^{2}}$,
the wavefunction components have the form ${r^{\gamma+1/2}}$, with ${\gamma=\sqrt{j^{2}-g^{2}}}$. For ${g>g_{c}\equiv|j|}$, the parameter ${\gamma}$ becomes imaginary, and the wavefunction oscillates dramatically towards the center.
We want to point out this remarkable behavior of the wavefunctions at short distance does not depend on the existence of magnetic field, because magnetic field related potential term has higher order of ${r}$ dependence than other terms of the potentials in Eqs. (8), (9), and is negligible when ${r\rightarrow 0}$. The above phenomenon is simply the supercritical instability, which is already well known in graphene Coulomb impurity problem.\cite{Pereira,Shytov,GamayunII} For the impurity problem, such instability signals the breakdown of the Dirac vacuum. Virtual electron-hole pairs are created, with negatively charged electrons going to infinity while the holes are bound to the Coulomb center (our impurity has negative charge). For the same problem under magnetic field, the virtual electrons can not go to infinity, because the effective potential is infinite when ${r\rightarrow\infty}$. When the supercritical instability happens, each Laudau level mixes with the quasistationary levels to better shield the large impurity charge, and we can {\em not} truncate the whole Hilbert space into one Landau level.

\subsection{WKB method}

WKB method is one of the basic and frequently-used methods to solve quantum mechanics problems without analytic solutions. Unlike perturbation theory, WKB method is not connected with the smallness of potential and thus has wider applicability range allowing one to study the qualitative behavior of the system. It also gives implicit or even explicit solutions for the energies as functions of parameters of the system, through which we can judge if the potential can be considered as perturbation from a semiclassical view. WKB method was originally created to approximately solve one dimensional, or radial part of higher dimensional Schrodinger particle problems. We formalize the WKB method for 2D massless Dirac particle problem below. Coulomb potential and uniform magnetic field are considered for our interest, but they can be replaced by any scalar and vector potential for general consideration.

Writing
\begin{equation}
\Phi(r)=\left(\begin{array}{c}
F(r)\\
G(r)\end{array}\right),
\end{equation}
the radial Eq. (3) becomes

\begin{equation}\Phi'(r)=\frac{1}{\hbar}D\Phi(r),\end{equation}
where
\begin{equation}\begin{split}D&\equiv \left(\begin{array}{cc}
\frac{\hbar j}{r}-\frac{\hbar}{2l_{B}^{2}}r & -(\frac{\hbar\epsilon}{l_{B}}-\frac{\hbar g}{r})\\
\frac{\hbar\epsilon}{l_{B}}-\frac{\hbar g}{r} & -(\frac{\hbar j}{r}-\frac{\hbar}{2l_{B}^{2}}r)\end{array}\right)\\
&=\left(\begin{array}{cc}
\frac{J}{r}-\frac{eB}{2c}r & -(\frac{E}{v_{F}}-\frac{Ze^{2}}{\kappa v_{F}r})\\
\frac{E}{v_{F}}-\frac{Ze^{2}}{\kappa v_{F}r} & -(\frac{J}{r}-\frac{eB}{2c}r)\end{array}\right),
\end{split}
\end{equation}
with the total angular momentum ${J=\hbar j}$.
Within WKB, we expand the solution of Eq. (11) in the form\cite{Rubinow,Kormanyos}
\begin{equation}\Phi(r)=e^{iy(r)/\hbar}\overset{\infty}{\underset{n=0}{\sum}}(-i\hbar)^{n}\varphi^{(n)}(r),\end{equation}
where ${y(r)}$ is a scalar function and ${\varphi^{(n)}(r)}$ are spinor functions.

For matrix ${D}$ in (12), the total angular momentum ${J}$ and energy ${E}$ are two conserved physical quantities, which are independent of ${\hbar}$. This may lead to some confusion because, say, ${J}$ equals to ${\hbar j}$ in quantum mechanical treatment of the system. However,  when we make ${\hbar\rightarrow0}$ and the theory returns to classical mechanics, quantum number ${j\sim1/\hbar}$ keeping the physical quantity invariant. Therefore, the matrix ${D}$ is independent of ${\hbar}$.

Inserting (13) into (11) and equating the coefficients of equal powers of ${\hbar}$, the first two equations of this set are
\begin{equation}
iy'(r)\varphi^{(0)}(r)=D\varphi^{(0)}(r),
\end{equation}
\begin{equation}
i\varphi^{(0)\prime}(r)+iy'(r)\varphi^{(1)}(r)=D\varphi^{(1)}(r).
\end{equation}
${iy'(r)}$ and ${\varphi^{(0)}(r)}$ are obtained as the eigenvalues and eigenvectors of matrix ${D}$:
\begin{equation}
\begin{split}
iy'(r)&\equiv \hbar \lambda_{i}(r)=\pm i\hbar p(r), \\
p(r)&=\sqrt{(\frac{\epsilon}{l_{B}}-\frac{g}{r})^{2}-(\frac{j}{r}-\frac{1}{2l_{B}^{2}}r)^2},
\end{split}
\end{equation}
\begin{equation}\varphi^{(0)}(r)\equiv\varphi_{i}(r)=Af_{i}(r)\left(\begin{array}{c}
\sqrt{sgn(S(r))(S(r)+\lambda_{i})}\\
s\sqrt{sgn(S(r))(S(r)-\lambda_{i})}\end{array}\right),\end{equation}
where subscript ${i=\pm}$ represents the two eigenvalues and their corresponding eigenvectors, ${S(r)\equiv\frac{j}{r}-\frac{1}{2l_{B}^{2}}r}$, ${s\equiv sgn(\frac{\epsilon}{l_{B}}-\frac{g}{r})\cdot sgn(\frac{j}{r}-\frac{1}{2l_{B}^{2}}r)}$, ${A}$ is any constant, ${f_{i}(r)}$ is the ${r}$ dependent common factor which has not been determined yet. For a complex number ${z}$ in this paper, we choose ${arg(z)}$ in the region ${(-\pi,\pi]}$, and ${arg(z^{1/2})}$=arg(z)/2. For this reason, ${sgn(S(r))}$ inside the square roots of Eq. (17) can not be factored out in order to keep the phase difference of the wavefunctions in two sublattices. Since matrix ${D}$ is not symmetric, left eigenvector ${\tilde{\varphi}^{(0)}(r)}$ satisfying ${\tilde{\varphi}^{(0)}(r)iy'(r)=\tilde{\varphi}^{(0)}(r)D}$ is introduced:
\begin{equation}
\begin{split}
&\tilde{\varphi}^{(0)}(r)\equiv\tilde{\varphi_{i}}(r)=Bg_{i}(r)\times\\
&\left(\begin{array}{cc}
\sqrt{sgn(S(r))(S(r)+\lambda_{i})}, & -s\sqrt{sgn(S(r))(S(r)-\lambda_{i})}\end{array}\right),
\end{split}
\end{equation}
where ${B}$ is any constant, ${g_{i}(r)}$ is any ${r}$ dependent common factors, and they are not important for our WKB results.
Multiplying Eq. (15) by ${\tilde{\varphi}^{(0)}(r)}$ on the left helps us to cancel the ${\varphi^{(1)}(r)}$ depended terms. Then Eq. (15) becomes
\begin{equation}
\tilde{\varphi}^{(0)}(r)\varphi^{(0)\prime}(r)=0.
\end{equation}
Substituting (17), (18) into Eq. (19), we obtain
\begin{equation}f_{i}(r)=\lambda_{i}(r)^{-\frac{1}{2}}.\end{equation}
In WKB approximation, we only keep functions ${y(r)}$ and ${\varphi^{(0)}(r)}$ in (13). The WKB approximate solution of Eq. (11) is obtained as
\begin{equation}
\Phi_{i}(r)=C\lambda_{i}^{-\frac{1}{2}}e^{\int^{r}\lambda_{i}dr}\left(\begin{array}{c}
\sqrt{sgn(S(r))(S(r)+\lambda_{i})}\\
s\sqrt{sgn(S(r))(S(r)-\lambda_{i})}\end{array}\right),
\end{equation}
where ${C}$ is a constant. The general solution ${\Phi(r)}$ could be written as the linear combination of ${\Phi_{+}(r)}$ and ${\Phi_{-}(r)}$:
\begin{widetext}
\begin{equation}
\Phi(r)=c_{1}\Phi_{+}+c_{2}\Phi_{-}=c_{1}p^{-\frac{1}{2}}e^{-i\int^{r}pdr}\left(\begin{array}{c}
\sqrt{sgn(S)(S-ip)}\\
s\sqrt{sgn(S)(S+ip)}\end{array}\right)+c_{2}p^{-\frac{1}{2}}e^{i\int^{r}pdr}\left(\begin{array}{c}
\sqrt{sgn(S)(S+ip)}\\
s\sqrt{sgn(S)(S-ip)}\end{array}\right),
\end{equation}
\end{widetext}
\normalsize
where ${r}$ is redefined as a dimensionless number representing the ratio of the distance from origin to magnetic length ${l_{B}}$, also redefine ${S(r)=j/r-r/2}$, ${p(r)=\sqrt{(\epsilon-g/r)^{2}-(j/r-r/2)^{2}}}$ and ${s\equiv sgn(\epsilon-g/r)\cdot sgn(j/r-r/2)}$ using the new dimensionless ${r}$, ${c_{1,2}}$ are constants fixed by boundary condition and normalization.

\begin{figure}[h]
\includegraphics[width=3.3in]{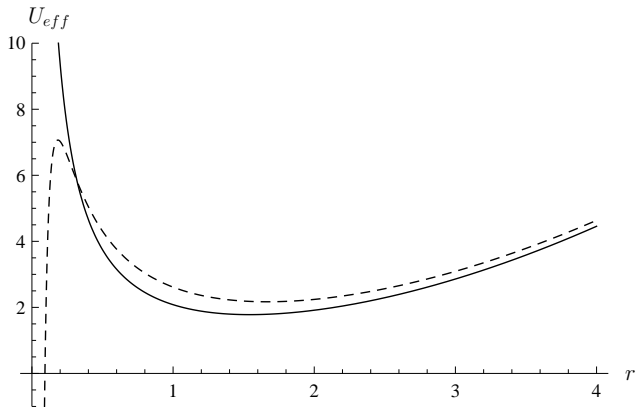}
\caption{WKB effective potentials ${U_{eff}\equiv(j^{2}-g^{2})/r^{2}+}$ ${2\epsilon g/r+r^{2}/4}$ for ${j=1/2}$. The solid line is for subcritical value ${g=0.49}$; the dashed line is for supercritical value ${g=0.7}$ (${g_{c}=0.5}$ when ${j=1/2}$). The energy ${\epsilon}$ is chosen to be 1.729, which is approximately the WKB energy of the 1st Landau level when ${g=0.49}$ calculated in Sec. III.}  \label{}
\end{figure}

To further obtain the BS condition for eigenenergies, we need to distinguish between classically allowed and forbidden regions. Defining WKB effective potential and WKB effective energy
\begin{equation}{U_{eff}\equiv(j^{2}-g^{2})/r^{2}+2\epsilon g/r+r^{2}/4,}\end{equation}
\begin{equation}{\epsilon_{eff}\equiv \epsilon ^2+j,}\end{equation}
the WKB wave number can be written as
\begin{equation}{p=\sqrt{\epsilon_{eff}-U_{eff}}.}\end{equation}
${U_{eff}}$ captures quantitatively the behavior of the supercritical instability, which is originally reflected by the wavefunction limiting behavior ${r^{\gamma+1/2}}$ at ${r \rightarrow 0}$ when the Schrodinger-like equations (7) are considered above. As shown in Fig. 1, for ${0\leq g<g_{c}=|j|}$, the WKB effective potential ${U_{eff}}$ is positive infinite at both ${r\rightarrow0}$ and ${r\rightarrow\infty}$, which allows us to use BS condition to obtain quantized energy levels and the WKB wavefunction vanishes at ${r \rightarrow 0}$. For ${g>g_{c}=|j|}$, ${U_{eff}}$ is still positively infinite at ${r\rightarrow\infty}$ but negatively infinite at ${r\rightarrow0}$, so the WKB wavefunction will oscillate as ${r\rightarrow 0}$. Just like what we can see from Eqs. (7), (8) and (9), the original Landau level states mix with the quasistationary states near the origin, and get the chance to be closer to the impurity to screen the impurity charge when the supercritical instability happens. For the following WKB calculation, we will only consider the weak coupling region (${g<g_{c}}$) and address the question: When are the electron states perturbatively connected to the states in a single Landau level?


There are four solutions for the quartic equation ${p^2(r)=0}$: ${r=\epsilon +\sqrt{\epsilon ^2-2 g+2 j}}$, ${-\epsilon +\sqrt{\epsilon ^2+2 g+2 j}}$, ${\epsilon -\sqrt{\epsilon ^2-2 g+2 j}}$ and ${-\epsilon -\sqrt{\epsilon ^2+2 g+2 j}}$. In the weak coupling region ${0\leq g<g_{c}=|j|}$, they are all real numbers. Two of the four real solutions are positive while the other two are negative. The two negative solutions have no physical meaning, but we want to keep them for the calculation in Sec. III. We can label the four solutions ${a}$, ${b}$, ${c}$, ${d}$ and require ${a>b>0>c>d}$. Using Zwaan's method (in Appendix), BS condition is obtained as

\begin{equation}\int_{b}^{a}p(r)dr=(n_{BS}+\frac{1-\theta(j)}{2})\pi,\end{equation}
where ${n_{BS}=0, 1, 2,...}$, ${\theta(j)}$ is step function, ${\theta(j)=1}$ for ${j>0}$ and ${\theta(j)=0}$ for ${j<0}$. For the special case of ${n_{BS}=0}$, classically allowed region disappears; this corresponds to the zeroth Landau level and will be discussed in detail later. The BS condition of the same case for Schrodinger particle was obtained in Ref. \onlinecite{Klama}. Besides the different forms of ${p(r)}$, Dirac particle BS condition has additional terms ${\pi/2-\theta(j)\pi/2}$, instead of ${\pi/2}$.

\begin{figure}[h]
\includegraphics[width=7.8cm]{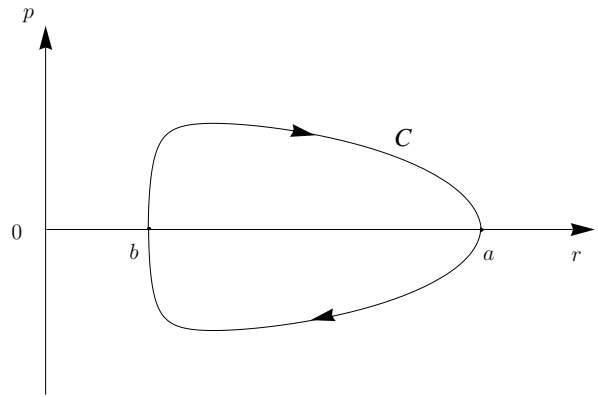}
\caption{Schematic diagram of the two branches of momenta ${\pm p}$ in phase space. The two branches merge continuously at the turning points ${a}$, ${b}$.}\label{}
\end{figure}
Compared to Zwaan's method used in Appendix, there is a more elegant way \cite{EBK,Kormanyos} to deduce the Bohr-Sommerfeld condition (26) without connecting boundary conditions. Moreover, this method helps us to see the origin of ${\theta(j)}$ in (26) straightforwardly.
In Fig. 2, we draw the two branches of momenta ${\pm p}$ as functions of $r$. They join at the turning points ${a}$, ${b}$ to form a single clockwise closed curve ${C}$ in phase space. Considering any term of the WKB wavefunction (22), say ${\Phi_{-}(r)}$, it must be single-valued after a full cycle along ${C}$.
First consider the exponent containing ${\int^{r}pdr}$. In one complete cycle, the phase change is ${\oint_{c}p(r)dr}$. There are additional phases introduced at each turning point by the amplitude factor ${p^{-\frac{1}{2}}}$. At each turning point, ${p}$ will change sign, which equals to adding a phase ${\pi}$ since ${exp(i\pi)=-1}$. Therefore ${\Phi_{-}(r)}$ gets an additional phase of ${-\pi/2}$ at each turning point.
Another kind of additional phases is introduced at each point where ${S(r)=0}$ by the spinor factor of ${\Phi_{-}(r)}$. At each such point ${sgn(S(r))ip(r)}$ inside the square root changes sign. Each ${S(r)=0}$ point gives an additional phase of ${\pi/2}$. Overall, the single-valuedness of the wavefunction demands
\begin{equation}
\oint_{c}p(r)dr-\mu\frac{\pi}{2}+\kappa\frac{\pi}{2}=n_{BS}2\pi,
\end{equation}
where ${\mu}$ is the number of turning points and ${\kappa}$ is the number of points where ${S(r)=0}$ in one complete cycle.
For negative ${j}$, ${S(r)=j/r-r/2}$ is always negative. For positive ${j}$, ${S(r)}$ is monotonically decreasing function, which equals to zero at one point in classically allowed region. Overall, for our case, ${\mu=2}$ and ${\kappa=2\theta(j)}$, so (27) returns to BS condition (26) directly.

\section{WKB results}
\subsection{WKB approximation without Coulomb impurity and the zeroth Landau level}

In this subsection, we first turn off the Coulomb potential, and consider the problem of one 2D Dirac particle in a perpendicular constant magnetic field. With exact solutions available, this problem enables us to compare WKB energies to exact solutions, and find the relationship between WKB quantum ${n_{BS}}$ and energy quantum ${n}$. In the units we have chosen, the exact eigenenergies of this problem are ${\epsilon=\pm\sqrt{2n}}$ where ${n=0}$, ${1}$, ${2}$, ${3}$, ${\ldots}$, and ${l\geq -n+1}$.

In the Bohr-Sommerfeld quantization condition (26), ${p(r)=\sqrt{\epsilon^{2}-(j/r-r/2)^{2}}}$ without Coulomb impurity and the integral can be carried out as
${\int_{b}^{a}p(r)dr=(\epsilon^{2}+j)\pi/2-|j|\pi/2.}$
For ${j>0}$, Eq. (26) gives ${\epsilon=\pm\sqrt{2n_{BS}}}$;
thus the WKB energies are identical to the exact energies! This also tells us ${n_{BS}}={n}$ is the Landau level index. For ${j<0}$, Eq. (26) gives ${\epsilon=\pm\sqrt{2(n_{BS}-l+1)}}$, which reproduces the Landau level spectrum when ${n_{BS}=n+l-1}$. Therefore, semiclassical energy correctly reproduces the Landau level spectrum.

From the argument above, we see WKB energy for the zeroth Landau level is obtained when ${n_{BS}=0}$. However, in this case  the two positive real roots of ${p^2(r)=0}$ equal to each other and classically allowed region becomes one point, thus the integral in Eq. (26) is zero. Such situation never occurs for Schrodinger particles as the presence of the $1/2$ shift in the BS condition. For Dirac particles such shift is zero for some cases. This is related to the $\pi$ Berry phase associated with their cyclotron motions (see Appendix of Ref. \onlinecite{kunreview} for a discussion of this point).

\begin{figure*}[t]
\includegraphics[width=7in]{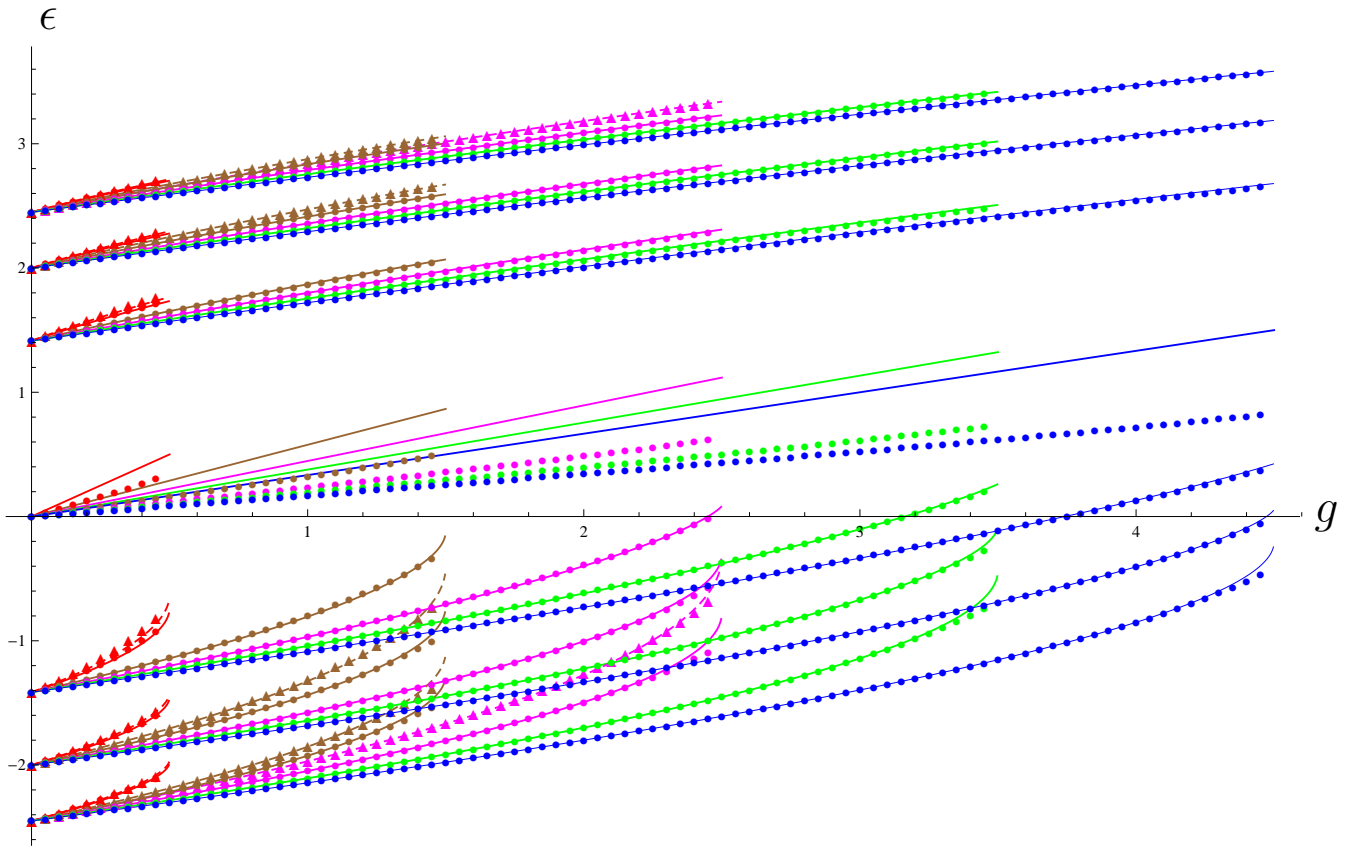}
\caption{(Color online) WKB energy and exact diagonalization (ED) energy levels from the negative 3rd to the positive 3rd Landau level and quantum number ${j}$ from its minimum value in each Landau level up to ${9/2}$. The lines are semiclassical energies. Negative ${j}$ are in dashed lines and positive ${j}$ are in regular lines. The circles (triangles) label the ED energies of positive (negative) ${j}$. For both WKB and ED energies, the spectra of ${|j|=1/2}$, ${3/2}$, ${5/2}$, ${7/2}$, ${9/2}$ are in red, brown, magenta, green, blue respectively. Energy spectrum with quantum number ${j}$ has the range ${g\in[0,|j|)}$.}  \label{}
\end{figure*}

\subsection{WKB approximation with Coulomb impurity}

When the Coulomb impurity is added (finite ${g}$), we can only carry out the energy calculation numerically. The integral in Eq. (26) can be expressed in terms of complete Legendre elliptic integrals of the first, ${F(\chi)}$, second, ${E(\chi)}$, and third, ${\Pi(\upsilon,\chi)}$ kinds.\cite{Lazur} The Bohr-Sommerfeld quantization condition (26) then gives the transcendental equation
\begin{equation}
\frac{1}{2}\left(-\mathcal{I}_{3}+\zeta\mathcal{I}_{1}-\beta \mathcal{I}_{0}-m^{2}\mathcal{I}_{-1}\right)=(n_{BS}+\frac{1-\theta(j)}{2})\pi,\end{equation}
with
${\zeta =4 j+4 \epsilon ^2,~\beta =8 g \epsilon,~m^2=4 j^2-4 g^2,}$
\small
\begin{equation}
\begin{split}
\mathcal{I}_{-1}&=\intop_{b}^{a}\frac{1}{R(r)}dr\\
&=\frac{2}{\sqrt{(a-c) (b-d)} b c} \left(b F(\chi )-(b-c) \Pi \left(\frac{c}{b} \nu ,\chi \right)\right),\\
\mathcal{I}_0~~&=\intop_{b}^{a}\frac{1}{R(r)}dr=\frac{2}{\sqrt{(a-c) (b-d)}} F(\chi ),\\
\mathcal{I}_1~~&=\intop_{b}^{a}\frac{r}{R(r)}dr=\frac{2}{\sqrt{(a-c) (b-d)}} (c F(\chi )+(b-c) \Pi (\nu ,\chi )),\\
T_2~~&=-\frac{1}{2 (1-\nu )} F(\chi )-\frac{\nu }{2 \left(\chi ^2-\nu \right) (1-\nu )}E(\chi )\\
&~~~+\frac{\chi ^2 (3-2 \nu )+\nu  (\nu -2)}{2 \left(\chi ^2 -\nu \right) (1-\nu )} \Pi (\nu ,\chi ),\\
\end{split}
\nonumber\end{equation}
\footnotesize
\begin{equation}
\begin{split}
T_3~~&=\left(\frac{\chi ^2}{4 \left(\chi ^2-\nu \right) (1-\nu )}-\frac{3 \left(\chi ^2 (3-2 \nu )+\nu  (\nu -2)\right)}{8 \left(\chi ^2-\nu \right) (1-\nu )^2}\right) F(\chi ),\\
&~~~+\left(\frac{3 \left(\chi ^2 (3-2 \nu )+\nu  (\nu -2)\right)^2}{8 \left(\chi ^2-\nu \right)^2 (1-\nu )^2}-\frac{3 \chi ^2-\nu  \left(1+\chi ^2\right)}{2 \left(\chi ^2-\nu \right) (1-\nu )} \right) \Pi (\nu ,\chi ),\\
&~~~-\frac{3 \nu  \left(\chi ^2 (3-2 \nu )+\nu  (\nu -2)\right)}{8 \left(\chi ^2-\nu \right)^2 (1-\nu )^2} E(\chi ),\\
\end{split}
\nonumber\end{equation}
\small
\begin{equation}
\begin{split}
\mathcal{I}_3~~&=\intop_{b}^{a}\frac{r^{3}}{R(r)}dr=\frac{2}{\sqrt{(a-c)(b-d)}}(c^3F(\chi)\\
&~~~+3c^2(b-c)\Pi(\nu,\chi)+3c(b-c)^2 T_2+(b-c)^3 T_3),
\end{split}
\nonumber\end{equation}
\normalsize
where
${\nu =(a-b)/(a-c)}$, ${\chi=\sqrt{\nu\text{ }(c-d)/(b-d)}}$ and ${R(r)=\sqrt{-\left(r^{4}-\zeta r^{2}+\beta r+m^{2}\right)}}$.

The semiclassical spectra as functions of ${g}$ for Landau levels from the ${-3}$rd to the ${3}$rd and ${j}$ up to ${9/2}$ are shown in Fig. 3. For ${g=0}$, the WKB solution coincides precisely with the exact solution, and the Landau level degeneracy of different angular momenta is lifted by finite ${g}$. For fixed ${g}$, the smaller ${|j|}$, the higher energy. The reason is that ${j\pm1/2}$ stands for the Dirac particle's orbital angular momentum in sublattice A(B). The particle with smaller angular momentum is closer to the impurity and feels stronger Coulomb potential.

To assess the accuracy of the semiclassical spectra, we use exact diagonalization (ED) method to obtain the energies of the same cases. For each total angular momentum labeled by ${j}$, the bases is from the ${-500}$th to the ${+500}$th Landau level's states when there is no impurity.  We observe that WKB energy levels are quite close to the ED energy levels in Fig. 3 except for the zeroth Landau level. First fix the value of impurity magnitude ${g}$. For smaller values of  ${|j|}$ and energy quantum ${|n|}$, the deviations from ED energy become larger. We make the assumption that ED results are accurate results, because the dimension of the bases is so large (1001D) and we are only considering the first few Landau levels. WKB method supposes the potential varies rather slowly in comparison to the de Broglie wavelength of the particle. In a certain Landau level (fixed n), the particle with smaller ${|j|}$ is closer to the impurity, and feels steeper potential, so the WKB method becomes more inaccurate. On the other hand, with certain ${j}$, the particles with smaller ${|n|}$ have less kinetic energy (larger de Broglie wavelength), then the WKB method becomes more inaccurate too. For ${g=0}$, WKB energy coincide with ED energy, with increasing ${g}$ and fixed quantum numbers ${j}$ and ${n}$, the difference becomes larger because potential becomes steeper. Since classically allowed region is only one point for the zeroth Landau level, wavelength becomes infinite and semiclassical approximation is not able to give accurate results.

\section{CONVERGENCE OF THE WKB SOLUTIONS}
In this section, we will consider the convergence of semiclassical solved energy in the region ${0\leq g<|j|}$. Based on the transcendental equations (26), (28) and analytic implicit function theorem,\cite{AIFT} we will argue that semiclassical energy levels do converge in the region ${0\leq g<|j|}$.

Write \small${f(g,\epsilon)\equiv\int_{b}^{a}p(r)dr=(-\mathcal{I}_{3}+\zeta\mathcal{I}_{1}-\beta \mathcal{I}_{0}-m^{2}\mathcal{I}_{-1})/2}$\normalsize. By analytic implicit function theorem, if we can prove for all the points satisfying ${g\in[0,|j|)}$ and ${\epsilon\neq g/\sqrt{2|j|}}$, function ${f(g,\epsilon)}$ is analytic and ${\partial f(g,\epsilon)/\partial\epsilon\neq0}$, then it can be concluded that there exists an explicit function ${\epsilon(g)}$ satisfying ${\{(g,\epsilon(g))|g\in[0,|j|)\}=\{(g,\epsilon)\in(0\leq g<|j|,\epsilon\neq g/\sqrt{2|j|})}$ ${|f(g,\epsilon)=[n_{BS}+(1-\theta(j))/2]\pi\}}$, and ${\epsilon(g)}$ is analytic in the region ${[0,|j|)}$. ${\epsilon= g/\sqrt{2|j|}}$ is a set of lines which make ${f(g,\epsilon)=0}$ (corresponding to the zeroth Landau level and being analytic obviously). The WKB results of nonzeroth Landau levels are definitely not in these lines.

It can be easily seen from Eq. (28) that ${f(g,\epsilon)}$ is analytic in the region ${(0\leq g<|j|,\epsilon\neq g/\sqrt{2|j|})}$, since ${a}$, ${b}$ are two positive numbers while ${c}$, ${d}$ are negative numbers. From the integral form of ${f(g,\epsilon)}$, we obtain ${\partial f(g,\epsilon)/\partial\epsilon=\int_{b}^{a}[(\epsilon-V)/p]dr}$. Because ${\epsilon-V\neq0}$ in the classically allowed region, ${\partial f(g,\epsilon)/\partial\epsilon\neq0}$.

Therefore, semiclassical energy functions ${\epsilon(g)}$ are analytic, then convergent in the region ${0\leq g<|j|}$.

\section{CONCLUSIONS}

In this paper we have used WKB approximation to study Coulomb impurity in the presence of a perpendicular uniform magnetic field in graphene. We find the solutions are smoothly (or perturbatively) connected to the states of isolated Landau level states when the impurity strength is below the supercritical instability critical point, thus lending support to the widely used Landau level projection approximation when treating many-electron problems in the quantum Hall regime. The WKB solutions are quantitatively accurate, except for the 0th Landau level states. On the other hand Landau level mixing becomes a {\em non-perturbative} effect beyond the supercritical instability critical point, signaling the breakdown of Landau level projection approximation.

\section*{Acknowledgments}
This work was supported by DOE grant No. DE-SC0002140 (YZ and KY), and the State of Florida (YB).

\section*{Appendix}
To obtain the Bohr-Sommerfeld quantization condition, we first need to connect the WKB wavefunctions in classically allowed and forbidden regions. In this work we use the method named after Zwaan (see Ref. \onlinecite{Landau} for an example), which is very instructive and does not make use of the exact solution (like Airy function). In Sec. II.A, we defined WKB effective potential (23), energy (24) and write ${p}$ in terms of them (25). The schematic diagram of ${U_{eff}}$ and ${\epsilon_{eff}}$ is plotted in Fig. 4.

Recall Eq. (22), the WKB wavefunctions in classically forbidden regions I and III are
\begin{equation}
\Phi_{I}(r)=Aq^{-\frac{1}{2}}e^{\int_{b}^{r}qdr}\left(\begin{array}{c}
\sqrt{sgn(S)(q+S)}\\
s\sqrt{sgn(S)(-q+S)}\end{array}\right),
\end{equation}
\begin{equation}
\Phi_{III}(r)=Cq^{-\frac{1}{2}}e^{-\int_{a}^{r}qdr}\left(\begin{array}{c}
\sqrt{sgn(S)(-q+S)}\\
s\sqrt{sgn(S)(q+S)}\end{array}\right).
\end{equation}
where ${q(r) \equiv ip(r)}$.
\begin{figure}[h]
\includegraphics[width=7.8cm]{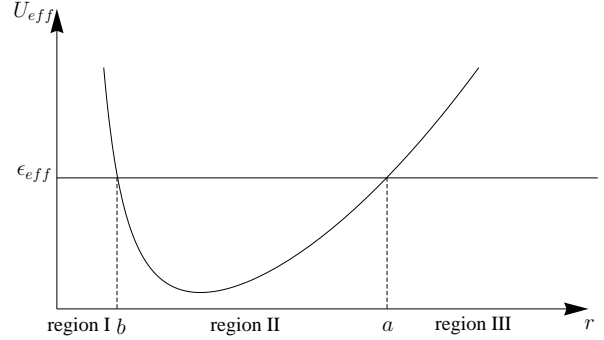}
\caption{Schematic diagram of WKB effective potential ${U_{eff}}$ and WKB effective energy ${\epsilon_{eff}}$ as functions of ${r}$.}\label{}
\end{figure}
And the WKB wavefunction in classically allowed region II can be written as
\begin{widetext}
\begin{equation}
\Phi_{II}(r)=B_{1}(-ip)^{-\frac{1}{2}}e^{-i\int_{b}^{r}pdr}\left(\begin{array}{c}
\sqrt{sgn(S)(-ip+S)}\\
s\sqrt{sgn(S)(ip+S)}\end{array}\right)+B_{2}(-ip)^{-\frac{1}{2}}e^{i\int_{b}^{r}pdr}\left(\begin{array}{c}
\sqrt{sgn(S)(ip+S)}\\
s\sqrt{sgn(S)(-ip+S)}\end{array}\right),
\end{equation}
\end{widetext}
or
\begin{widetext}
\begin{equation}
\Phi_{II}(r)=B'_{1}(-ip)^{-\frac{1}{2}}e^{-i\int_{a}^{r}pdr}\left(\begin{array}{c}
\sqrt{sgn(S)(-ip+S)}\\
s\sqrt{sgn(S)(ip+S)}\end{array}\right)+B'_{2}(-ip)^{-\frac{1}{2}}e^{i\int_{a}^{r}pdr}\left(\begin{array}{c}
\sqrt{sgn(S)(ip+S)}\\
s\sqrt{sgn(S)(-ip+S)}\end{array}\right).
\end{equation}
\end{widetext}
The above two forms of the WKB wavefunctions in the classically allowed region differ due to a different choice of the lower limits in the integral, which correspond to the two turning points. This is done in order to match the wavefunctions in the classically forbidden regions I and III.

Before connecting the wavefunctions of different regions, we need to have a mathematical interlude. Write
\begin{equation}
\begin{split}
\sqrt{\frac{ip+S}{\epsilon-V}}=e^{-i \int_{a,b}^{r}\frac{1}{2p}(\frac{V'S}{\epsilon-V}+S')dr+ic_{a,b}};\\
\sqrt{\frac{-ip+S}{\epsilon-V}}=e^{i \int_{a,b}^{r}\frac{1}{2p}(\frac{V'S}{\epsilon-V}+S')dr-ic_{a,b}},
\end{split}
\end{equation}
where the two equations are complex conjugate to each other, ${V(r)=g/r}$ for our case. Constants ${c_{a,b}}$ correspond to the two different lower limits ${a}$ and ${b}$ of each integral. It is easy to check that these constants should be imaginary so we write them as ${ic_{a,b}}$ where ${c_{a,b}}$ are real numbers. The relations between the square roots in the wavefunctions (29-32) to the ones in (33) are \begin{equation}{\sqrt{sgn(S)(ip+S)}=e^{-it}\sqrt{|\epsilon-V|}\sqrt{\frac{ip+S}{\epsilon-V}}},\end{equation} \begin{equation}{\sqrt{sgn(S)(-ip+S)}=e^{it}\sqrt{|\epsilon-V|}\sqrt{\frac{-ip+S}{\epsilon-V}}},\end{equation} where ${t=[sgn(\epsilon-V)-sgn(S)]\pi/4}$.

Now, we begin to use Zwaan's method to connect the wavefunctions in region II and III. Near ${r=a}$, we can make the linear approximation ${q=\sqrt{|F_{0}|(r-a)}}$, where ${F_{0}=\partial[(\epsilon-V(r))^{2}-S(r)^{2}]/\partial r|_{r=a}}$. Write everything in complex plane
\begin{figure}[h]
\includegraphics[width=8cm]{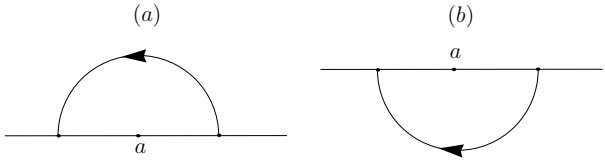}
\caption{Two different paths of WKB wavefunction passes from region III (classically forbidden region) to region II (classically allowed region) in complex plane.}\label{}
\end{figure}
\begin{equation}r-a=\rho e^{i\phi}, ~\intop_{a}^{r}\sqrt{r-a}dr=\frac{2}{3}\rho^{\frac{3}{2}}(cos\frac{3}{2}\phi+isin\frac{3}{2}\phi).\end{equation}
When region III and II 's wavefunctions are connected through the upper semicircle path as Fig. 5(a)
\begin{equation}q(r)\rightarrow ip(r), ~-\int_{a}^{r}q(r)dr\rightarrow-i\int_{a}^{r}p(r)dr,\end{equation}
\small
\begin{equation}
\begin{split}
\Phi_{III}(r)&=Cq^{-\frac{1}{2}}e^{-\int_{a}^{r}qdr}\left(\begin{array}{c}
\sqrt{sgn(S_{a})(-q+S)}\\
s\sqrt{sgn(S_{a})(q+S)}\end{array}\right)\\
&\rightarrow C(ip)^{-\frac{1}{2}}e^{-i\int_{a}^{r}pdr}\left(\begin{array}{c}
\sqrt{sgn(S_{a})(-ip+S)}\\
s\sqrt{sgn(S_{a})(ip+S)}\end{array}\right)\\
&=C\sqrt{|\epsilon-V|}(ip)^{-\frac{1}{2}}e^{-i\int_{a}^{r}pdr}\left(\begin{array}{c}
e^{it_{a}}\sqrt{\frac{-ip+S}{\epsilon-V}}\\
se^{-it_{a}}\sqrt{\frac{ip+S}{\epsilon-V}}\end{array}\right),
\end{split}
\end{equation}
\normalsize
where ${S_{a,b}=S(r=a,b)}$.
When we connect region III and II's wavefunctions through the lower semicircle path as Fig. 5(b),
\begin{equation}q(r)\rightarrow-ip(r), ~-\int_{a}^{r}q(r)dr\rightarrow i\int_{a}^{r}p(r)dr,\end{equation}
\small
\begin{equation}
\begin{split}
\Phi_{III}(r)&=Cq^{-\frac{1}{2}}e^{-\int_{a}^{r}qdr}\left(\begin{array}{c}
\sqrt{sgn(S_{a})(-q+S)}\\
s\sqrt{sgn(S_{a})(q+S)}\end{array}\right)\\
&\rightarrow C(-ip)^{-\frac{1}{2}}e^{i\int_{a}^{r}pdr}\left(\begin{array}{c}
\sqrt{sgn(S_{a})(ip+S)}\\
s\sqrt{sgn(S_{a})(-ip+S)}\end{array}\right)\\
&=C\sqrt{|\epsilon-V|}(-ip)^{-\frac{1}{2}}e^{i\int_{a}^{r}pdr}\left(\begin{array}{c}
e^{-it_{a}}\sqrt{\frac{ip+S}{\epsilon-V}}\\
se^{it_{a}}\sqrt{\frac{-ip+S}{\epsilon-V}}\end{array}\right),
\end{split}\end{equation}
\normalsize
where ${t_{a}=[sgn(\epsilon-V_{a})-sgn(S_{a})]\pi/4}$ and ${V_{a,b}=V(r=a,b)}$. Let ${B'_{1}=e^{-i\pi/2}C}$, ${B'_{2}=C}$, Eq. (29) becomes
\small
\begin{equation}\begin{split}
&\Phi_{II}(r)=C(\frac{|\epsilon-V|}{p})^{\frac{1}{2}}\\
&\times\left(\begin{array}{c}
cos[\int_{a}^{r}[p-\frac{1}{2p}(\frac{V'S}{\epsilon-V}+S')]dr+c_{a}-t_{a}+\frac{\pi}{4}]\\
sgn(\epsilon-\frac{g}{r}) cos[\int_{a}^{r}[p+\frac{1}{2p}(\frac{V'S}{\epsilon-V}+S')]dr-c_{a}+t_{a}+\frac{\pi}{4}]\end{array}\right).
\end{split}\end{equation}
\normalsize
Similarly, by connecting the wavefunctions of region I and II, we obtain
\small
\begin{equation}\begin{split}&\Phi_{II}(r)=A(\frac{|\epsilon-V|}{p})^{\frac{1}{2}}\\
& \times\left(\begin{array}{c}
cos[\int_{b}^{r}(p-\frac{1}{2p}(\frac{V'S}{\epsilon-V}+S'))dr+c_{b}-t_{b}-\frac{\pi}{4}]\\
sgn(\epsilon-\frac{g}{r}) cos[\int_{b}^{r}(p+\frac{1}{2p}(\frac{V'S}{\epsilon-V}+S'))dr-c_{b}+t_{b}-\frac{\pi}{4}]\end{array}\right),\end{split} \end{equation}
\normalsize
where ${t_{b}=[sgn(\epsilon-V_{b})-sgn(S_{b})]\pi/4}$.

Eq. (41), (42) then give us the Bohr-Sommerfeld quantization condition
\begin{equation}\int_{b}^{a}[p-\frac{1}{2p}(\frac{V'S}{\epsilon-V}+S')]dr-c_{a}+c_{b}+t_{a}-t_{b}=(n_{BS}+\frac{1}{2})\pi.\end{equation}
where ${n_{BS}=0, 1, 2,...}$, ${p(r)\equiv-iq(r)=[(\epsilon-g/r)^{2}-(j/r}$ ${-r/2)^{2}]^{1/2}}$. In region II where ${p^{2}(r)=(\epsilon-V)^{2}-S^{2}=}$ ${(\epsilon-V)^{2}-(j/r}$ ${-r/2)^{2}>0}$, ${\epsilon-V(r)}$ can not be zero, so the signs of ${\epsilon-V(r)}$ at point ${r=a}$ and ${b}$ are the same. It is also easy to see, for ${j<0}$ ${S(r)}$ is always negative; for ${j>0}$, ${S_{b}}$ is positive and ${S_{a}}$ is negative. Overall, we obtain  ${t_{a}-t_{b}=\theta(j)\pi/2}$, where ${\theta(j)=1}$ for ${j>0}$ and ${\theta(j)=0}$ for ${j<0}$. Analyzing Eq. (33) by substitute ${r=a}$ and ${b}$, we can always get ${-\int_{b}^{a}[V'S/(\epsilon-V)+S']/2pdr-c_{a}+c_{b}=0}$. Finally, Eq. (43) is simplified as
\begin{equation}\int_{b}^{a}p(r)dr=(n_{BS}+\frac{1-\theta(j)}{2})\pi.\end{equation}

\end{document}